\documentclass[journal=inoraj,manuscript=article,layout=twocolumn]{achemso}

\setkeys{acs}{articletitle = true}
\setkeys{acs}{email=false}

\usepackage{charter}
\usepackage{color}
\usepackage{mathtools}
\usepackage{array}

\newcommand*{\nsii}{(Nb$_4$Se$_{15}$I$_2$)I$_2$}

\author{Kejian Qu}
\affiliation{Department of Physics and Materials Research Laboratory, University of Illinois at Urbana-Champaign, Urbana, IL 61801, USA}

\author{Zachary W. Riedel}
\affiliation{Department of Materials Science and Engineering and Materials Research Laboratory, University of Illinois at Urbana-Champaign, Urbana, IL 61801, USA}

\author{Iri\'{a}n S\'{a}nchez-Ram\'{i}rez}
\affiliation{Donostia International Physics Center, P. Manuel de Lardizabal 4, 20018 Donostia-San Sebastian, Spain}

\author{Simon Bettler}
\affiliation{Department of Physics and Materials Research Laboratory, University of Illinois at Urbana-Champaign, Urbana, IL 61801, USA}

\author{Junseok Oh}
\affiliation{Department of Physics and Materials Research Laboratory, University of Illinois at Urbana-Champaign, Urbana, IL 61801, USA}

\author{Emily N. Waite}
\affiliation{Department of Physics and Materials Research Laboratory, University of Illinois at Urbana-Champaign, Urbana, IL 61801, USA}

\author{Nadya Mason}
\affiliation{Department of Physics and Materials Research Laboratory, University of Illinois at Urbana-Champaign, Urbana, IL 61801, USA}

\author{Peter Abbamonte}
\affiliation{Department of Physics and Materials Research Laboratory, University of Illinois at Urbana-Champaign, Urbana, IL 61801, USA}

\author{Fernando de Juan Sanz}
\affiliation{Donostia International Physics Center, P. Manuel de Lardizabal 4, 20018 Donostia-San Sebastian, Spain}
\alsoaffiliation{Ikerbasque, Basque Foundation for Science, 48013 Bilbao, Spain}

\author{Maia G. Vergniory}
\affiliation{Donostia International Physics Center, P. Manuel de Lardizabal 4, 20018 Donostia-San Sebastian, Spain}
\alsoaffiliation{The Max Planck Institute for Chemical Physics of Solids, 01187 Dresden, Germany}

\author{Daniel P. Shoemaker}\email{dpshoema@illinois.edu}
\affiliation{Department of Materials Science and Engineering and Materials Research Laboratory, University of Illinois at Urbana-Champaign, Urbana, IL 61801, USA}
\title[Short title]
{A new quasi-one-dimensional transition metal chalcogenide semiconductor (Nb$_4$Se$_{15}$I$_2$)I$_2$}


\begin{document}


\begin{abstract}
The discovery of new low-dimensional transition metal chalcogenides is contributing to the already prosperous family of these materials.
In this study, needle-shaped single crystals of a new quasi-one-dimensional material {\nsii} were grown by chemical vapor transport, and the structure was solved by single crystal X-ray diffraction (XRD). The new structure has one-dimensional (Nb$_4$Se$_{15}$I$_2$)$_n$ chains along the [101] direction, with two I$^-$ ions per formula unit directly bonded to Nb$^{5+}$. The other two I$^-$ ions are loosely coordinated and intercalate between the chains. 
Individual chains are chiral, and stack along the $b$ axis in opposing directions, giving space group $P2_1/c$.
The phase purity and crystal structure was verified by powder XRD.
Density functional theory calculations show {\nsii} to be a semiconductor with a direct band gap of around 0.6~eV. Resistivity measurements of bulk crystals and micro-patterned devices demonstrate that {\nsii} has an activation energy of around 0.1~eV, and no anomaly or transition was seen upon cooling. {\nsii} does not undergo structural phase transformation from room temperature down to 8.2~K, based on cryogenic temperature single crystal XRD.
This compound represents a well-characterized and valence-precise member of a diverse family of anisotropic transition metal chalcogenides.
\end{abstract}


\section{Introduction}
Low-dimensional transition metal chalcogenides have been under intensive study for decades, and formative works on the crystal chemistry,\cite{TMC_crystal_chemistry, TMC_CVT, ligand_field} band structure,\cite{Band_structure_Goodenough, EELS_band_structure} and structural phase transitions\cite{structural_phase_transition} were carried out from the 1960s. Some of these materials are now undergoing a renaissance due to renewed interest in charge density wave (CDW) formation\cite{STM_CDW} and topological phase transitions in Weyl semimetals.\cite{Weyl_semimetal_TMC}
The propensity for 1-D and 2-D exfoliation, the strong anisotropy, and their semiconducting behavior with few bands (and thus well-defined orbital contributions to transport) at the Fermi level are among the major merits of transition metal chalcogenides, which have been reviewed extensively by Manzeli, et al.\cite{TMDC_reivew} and Petra, et al.\cite{1D_TMTC_review}
Despite a sustained interest in this class of materials, the precise mechanisms that lead to their synthesis and transport are still under investigation, and accordingly the landscape of stability has not been fully explored.

CDW formation\cite{CDW_review} is one of the defining hallmarks properties of some transitional metal chalcogenides. Quasi-one-dimensional materials such as NbSe$_3$\cite{NbSe3_first_structure, NbSe3_CDW} and TaS$_3$\cite{TaS3_first_structure, TaS3_CDW, TaS3_NbSe3_CDW} undergo CDW transitions, where a simple interpretation would predict a Peierls distortion along the 1D chains below their respective CDW transition temperatures. In reality, the band structures are complex and there are many ways to lower the electronic energies of these systems, so CDW wavevectors are often in different or multiple directions.
Low-temperature X-ray or electron diffraction is typically the most clear evidence of these CDW transitions and orderings: when superlattices form due to the onset of a CDW, satellite reflections around the main Bragg reflections are apparent, for example in NbSe$_3$ and TaS$_3$.\cite{NbSe3_CDW, TaS3_NbSe3_resistivity_diffraction} 
Another key signature of CDW transitions is the deviation from Arrhenius resistivity behavior, which is embodied by the non-linearity of the logarithmic resistivity ln ($\rho$) versus $1/T$, such as the steep increase in resistivity seen at 240~K for monoclinic TaS$_3$, and 145~K and 59~K in NbSe$_3$,\cite{TaS3_NbSe3_resistivity_diffraction, resistivity_TaS3} as temperature decreases.

Closely related to NbSe$_3$ and TaS$_3$, another group of ternary quasi-one-dimensional transition metal chalcogenides were discovered shortly after. These compounds have iodine intercalated between the chains,\cite{structure_whole_family} such as (TaSe$_4$)$_2$I (space group $I422$),\cite{TSI_first_structure} (NbSe$_4$)$_2$I (isostructural with (TaSe$_4$)$_2$I), (NbSe$_4$)$_3$I (space group $P4/mnc$),\cite{NS3I_first_structure} and (NbSe$_4$)$_{3.33}$I (space group $P4/mcc$).\cite{NS333I_structure}
These materials have (TaSe$_4$)$_n$ or (NbSe$_4$)$_n$ chains along the $c$-axis, with all Se ions dimerized into (Se$_2$)$^{2-}$ units and bonded to either Ta or Nb.
Ta and Nb ions have alternating 4+ and 5+ valence, leaving the (TaSe$_4$)$_n$ or (NbSe$_4$)$_n$ chain with partially-filled conducting $d$ bands.
(TaSe$_4$)$_2$I,\cite{TSI_first_CDW} (NbSe$_4$)$_2$I,\cite{NS2I_CDW} (NbSe$_4$)$_3$I,\cite{NS3I_low_T_structure} and (NbSe$_4$)$_{3.33}$I\cite{NS333I_CDW} undergo CDW transitions, and low-temperature X-ray diffraction (XRD) as well as resistivity measurements were used to demonstrate the onset of CDW.\cite{TSI_NS2I_XRD_resistivity_1, TSI_NS2I_XRD_resistivity_2}
Traditionally, single crystals of (TaSe$_4$)$_2$I and (NbSe$_4$)$_3$I have been successfully synthesized by chemical vapor transport (CVT)\cite{CVT_review} where intermediate gaseous species deposit the target compounds at the low-temperature side of the reaction vessel at temperatures ranging from 400$^\circ$C to 800$^\circ$C.\cite{TSI_Maki, structure_whole_family}
(TaSe$_4$)$_2$I is currently under renewed focus, particularly due to claims that it may harbor ``axion insulator'' behavior if the CDW wavevector is coincident with one of the Ta $d$-bands crossing the Fermi energy, the location of which is termed a Weyl point due to the chirality of the compound.\cite{TSI_nature, TSI_Nature_Physics} The band-crossing at the Fermi energy of (TaSe$_4$)$_2$I is a strong hint for the potential onset of CDW,\cite{TSI_DFT, TSI_ARPES, TSI_Fahad} though the CDW wavevector is not coincident with the crossing point.
Due to the complexities of these materials and the urgent need to compare and benchmark how charge transport phenomena correlate with band structures, it is of great interest to discover new transition metal chalcogenides with similar elemental compositions and a quasi-one-dimensional structure.

Here we present a new quasi-one-dimensional compound {\nsii}, with a structure related to (TaSe$_4$)$_2$I and (NbSe$_4$)$_3$I. 
Formed from analogous CVT reactions, \nsii\ is a band insulator with a single Se per formula unit acting as a divalent (non-dimerized) ion and two of the four $I^-$ ions present in the 1-D chain, hence the formula is deliberately written {\nsii}, instead of simply Nb$_4$Se$_{15}$I$_4$. The complex chain stoichiometry enforces valence-precise Nb$^{5+}$, and our transport, diffraction, and DFT examinations conclude that \nsii\ is a band insulator and has no CDW transition since no electronic bands cross the Fermi energy.

\section{Methods}

\subsection{Synthesis}

The CVT synthesis of {\nsii} gained its insight from the CVT synthesis of (TaSe$_4$)$_2$I,\cite{TSI_Maki} and the results depend heavily on details, including sample mass, stoichiometry, form (powder or wire) of ingredients, and temperature profile, etc.
It started with mixing elemental powders in a mortar with a ratio of Nb(99.8\%) : Se(99.999\%) : I(99.8\%) = 1 : 8 : 10.3. The total sample mass was about 0.9~g. After fully grinding with a pestle, the mixture tends to form a porous, loosely-bound agglomerate, and niobium powder will not stick to the surface of the mortar.
Then the powder mixture was loaded to a fused quartz tube, which was sealed under vacuum.
The sealed tube was heated up in a two-zone furnace with high and low temperature sides being 420$^\circ$C and 280$^\circ$C, respectively. The sample was slowly heated up to the target temperatures for 10 hours, and stayed for 90 hours before cooling down naturally.

After reaction, dense needle-shaped crystals formed on the inner wall of the tube, in a large region where the temperature ranges from 350$^\circ$C to 290$^\circ$C roughly.
The crystals did not form directly on the wall, but onto a layer of iodine (or iodine containing compound) which coated the lower-temperature half of the tube after reaction.
A picture of the tubes after reaction is shown in Figure S1 in Supporting Information.\cite{supplement}
The typical dimensions of these crystals are $5\times0.1\times0.1$~mm$^3$. Some crystals are ribbon-shaped, which has a thickness even smaller than 0.1~mm, and some largest ones can be around $10\times0.5\times0.5$~mm$^3$.
Loose powder of (NbSe$_4$)$_3$I formed as residue at the high-temperature side of the tube, as a byproduct.
Slightly different temperature profiles were attempted and {\nsii} crystals were able to grow as well, but the size and yield of the crystals were inferior to the aforementioned temperature profile, and the byproduct (NbSe$_4$)$_3$I had a much larger yield.
This further indicates the sensitivity of the reaction with the reaction conditions.

\subsection{Structure determination}

Single crystal XRD was performed on a Bruker D8 Venture Duo at room temperature and the structure of {\nsii} was solved.
Later on, powder XRD was performed on a Bruker D8 Advance using crushed crystals. The needle crystals cannot be ground into fine powder so hundreds of thin needles were crushed and treated as the powder sample. The result was refined by GSAS II\cite{GSAS} to verify the structure of {\nsii}.
Scanning electron microscopy was performed in a ThermoFisher Axia ChemiSEM.

\subsection{Band structure calculation}

First-principles density functional theory (DFT) calculations were performed using Vienna ab initio simulation package (VASP)\cite{VASP1, VASP2} with projector-augmented wave pseudo-potentials and two approximations: the Perdew Burke Ernzerhof parametrization (PBE)\cite{Perdew96} and the modified Becke-Johnson method (mBJ).\cite{Tran2009} The mBJ method was chosen in order to obtain a more accurate description for the band gap of {\nsii} and check the validity of PBE approximation for density of states (DOS) calculations. The electronic band structure was obtained employing experimental structural data refined by single crystal XRD, using a plane-wave basis with a kinetic cutoff of $520$~eV and a $7\times 3 \times 7$ $k$-mesh for both approximations. Orbital-resolved DOS results were obtained using a $11\times 5 \times 11$ $k$-mesh using the tetrahedron method with Blöchl corrections\cite{Blochl} and PBE approximation. All calculations were performed considering spin-orbit coupling (SOC).

\subsection{Resistivity measurements}

An in-line four-point resistivity measurement was carried out with Quantum Design Physical Property Measurement System (PPMS). A long, thick and clean single crystal was chosen and the contacts were made using gold wires and silver epoxy.
The crystal was approximately 0.5~mm wide and 0.5~mm thick, with two inner voltage leads having a distance of about 1~mm.

As a comparison, one of the crystals was also exfoliated using a scotch tape into flakes with thickness less than 100~nm. E-beam lithography was used to pattern the electrical contacts in four-point configuration on these devices, followed by deposition of Ti(5~nm)/Au(60~nm). A micrograph of the device is shown in Figure S4 in Supporting Information.\cite{supplement} The electric resistance of the device was measured in the PPMS and a custom setup including a Keithley 2400 multimeter.

\subsection{Phase transition determination}

Three-dimensional X-ray maps of momentum space were obtained using a Mo K$\alpha$ ($\lambda=0.7107$~{\AA}) microspot X-ray source and a Mar345 image plate detector by sweeping crystals through an angular range of 20$^\circ$ with an exposure time of 180 seconds per image. The sample was first cooled down to base temperature, at 8.2~K, and then warmed up. Diffraction data was collected at 8.2~K, 100~K, 200~K, 250~K, and 290~K. Figure S5 and Figure S6 in Supporting Information show the diffraction data at 290~K and 8.2~K, respectively, with diffraction spots indexed for 8.2~K.\cite{supplement}

Differential scanning calorimetry (DSC) was performed using TA Discovery 2500, cycling in the temperature range of -125$^\circ$C to 25$^\circ$C, to determine potential phase transition. The result is shown in Figure S7 in Supporting Information.\cite{supplement}

\section{Results and Discussion}

\subsection{Crystal structure of {\nsii}}

\begin{figure}
    \centering
    \includegraphics[width=0.95\columnwidth]{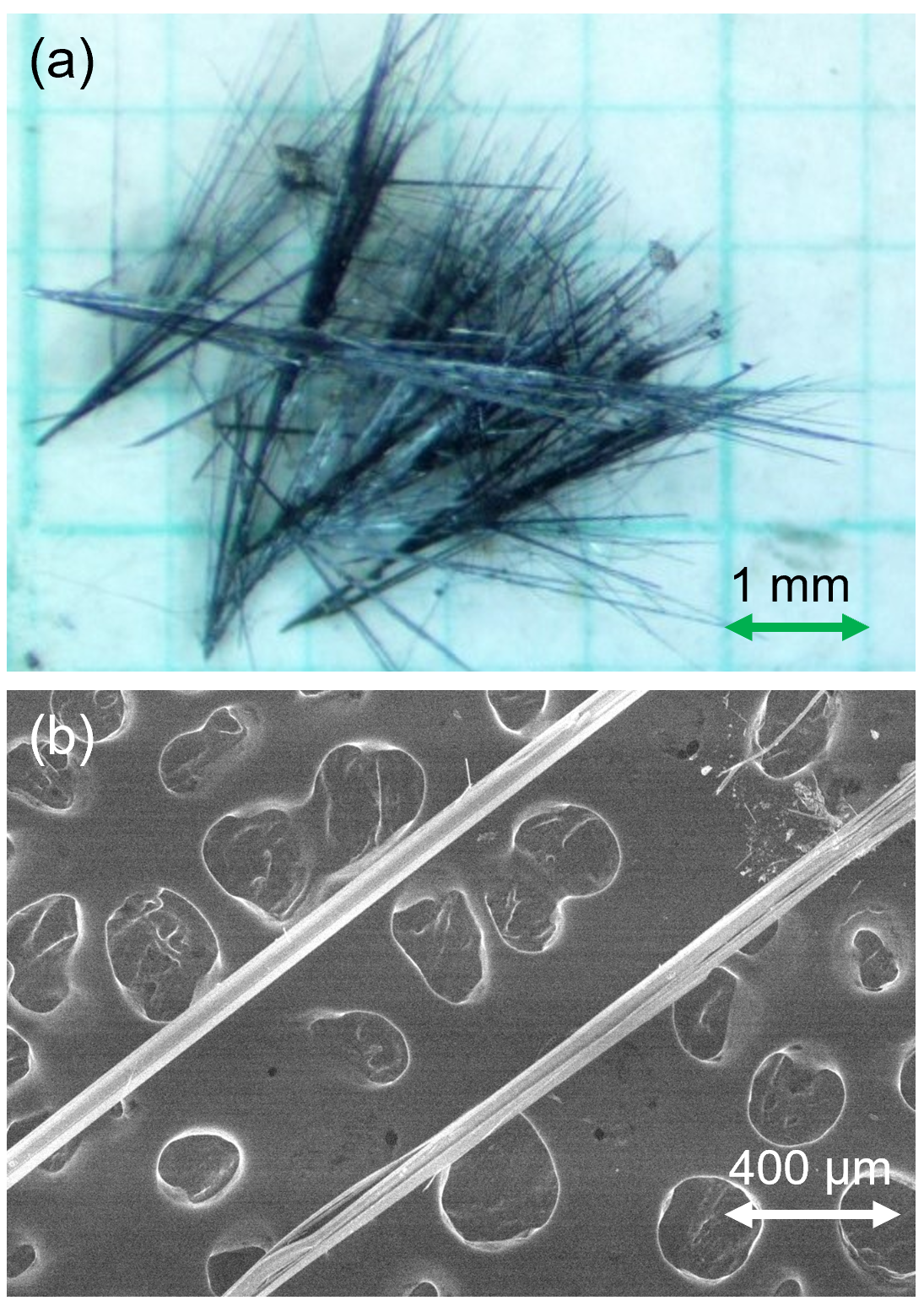}
    \caption{(a) Photograph of typical {\nsii} crystals obtained from CVT. (b) SEM of two {\nsii} crystal bundles on carbon tape.}
    \label{fig: pictures}
\end{figure}

Bundles of needle-shaped single crystals of {\nsii} grew from common nucleation centers, as shown in Figure \ref{fig: pictures}(a). These multi-millimeter single crystals can be further exfoliated into thinner strands easily. Figure \ref{fig: pictures}(b) is the SEM picture of two exfoliated, thinner needle crystals. The splitting of the strands is more clear in the lower crystal, suggesting weak van der Waals bonding between the chains.

\begin{table}
    \centering
    \caption{Overview of single-crystal XRD refinement of \nsii. Atomic positions are given in Table S1 in Supporting Information.}
    \begin{tabular}{c c}
    \hline
Space group & $P2_1/c$ \\
$a$	& 8.8807(2) \AA \\
$b$	& 25.8368(7) \AA \\
$c$ & 11.6222(3) \AA \\
$V$ & 2593.70(11) \AA$^3$ \\
$Z$ & 4 \\
Absorption coefficient & 27.527 \\
Experimental density & 5.285~g/cm$^3$ \\
$F$(000) & 3544 \\
Reflections & 7932\\
Goodness of fit & 1.097 \\
Number of patterns & 208 \\
$R_\mathrm{int}$ & 6.09~\% \\
$R_1$ & 2.47~\% \\
$wR_2$ & 7.10~\% \\
$I/\sigma$ & 40.2 \\
$d_\mathrm{min}$ & 0.70~\AA \\
    \hline
    \end{tabular}
    \label{tab: structure}
\end{table}

\begin{figure}
    \centering
    \includegraphics[width=\columnwidth]{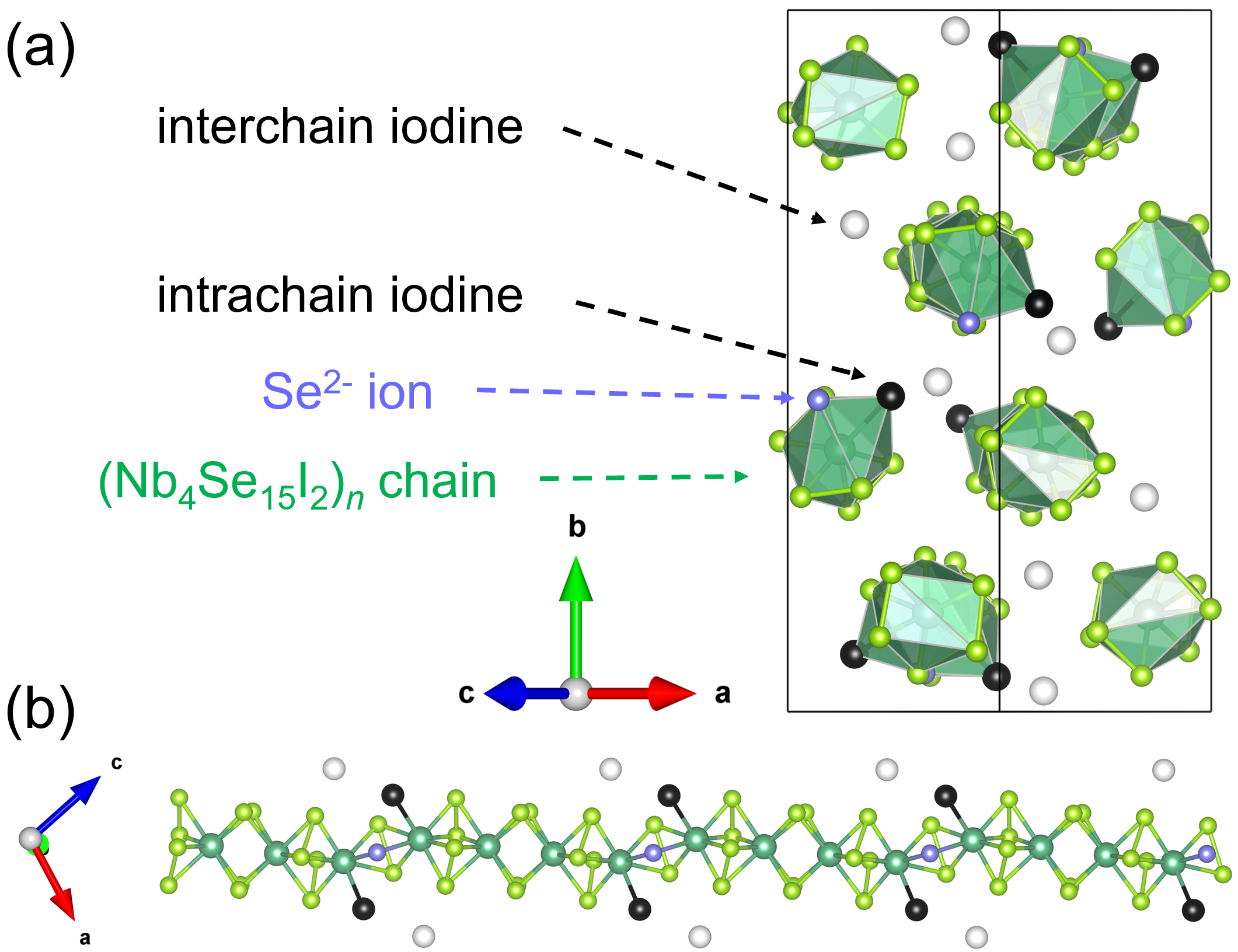}
    \caption{(a) Unit cell of {\nsii} viewed down 1-D chains showing the hexagonal packing of chains with interchain I$^-$ (white), intrachain I$^-$ (black), and a single Se$^{2-}$ (light blue), whereas all other selenium is dimerized as (Se$_2$)$^{2-}$ (light green). (b) View along $b$, normal to the 1-D chains, highlighting the ions and coordinations with the same color scheme.}
    \label{fig: structure}
\end{figure}

The crystal structure of {\nsii} was solved from single crystal XRD, with details shown in Table \ref{tab: structure} and atomic positions in Table S1.
The unit cell is shown in Figure \ref{fig: structure}(a), where (Nb$_4$Se$_{15}$I$_2$)$_n$ form one-dimensional chains along [101] direction (Figure \ref{fig: structure}(b)).
Half of the iodine ions (black, intrachain) are directly bonded to niobium ions, while the other half iodine ions (white, interchain) lie between the chains and are weakly associated, with the closest ion being Se at 3.1720(8)~{\AA}. This explains why the chemical formula is written {\nsii} intentionally, in analogy to (TaSe$_4$)$_2$I, to emphasize the fact that the iodine ions are split evenly into intra- and interchain positions.

\begin{figure*}
    \centering
    \includegraphics[width=\textwidth]{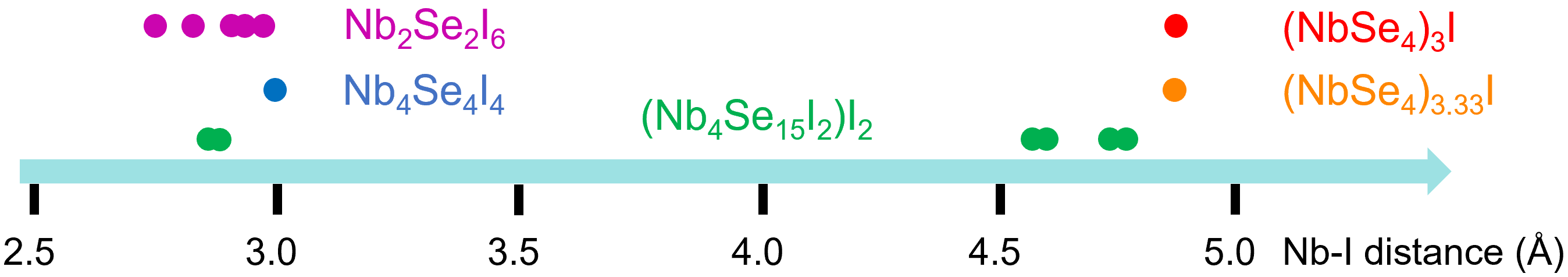}
    \caption{Nb--I distances of Nb--Se--I compounds, demonstrating the difference of intrachain Nb--I bonding (short) and interchain (long) Nb--I distances. Each material may have multiple Nb--I distances.}
    \label{fig: distance}
\end{figure*}

The distance between a niobium ion with the iodine directly bonded is either 2.8889(7)~{\AA} or 2.9155(7)~\AA\ in \nsii. On the contrary, the distance between a niobium ion with the closest interchain iodine has values of 4.5732(7)~\AA, 4.6011(8)~\AA, 4.7254(7)~\AA, and 4.7703(7)~\AA, depending on which of the four types of niobium ions is considered, since they have slightly different chemical environments.
As a comparison, the Nb--I bonding distances for other Nb--Se--I ternary compounds are similar. For compounds with Nb-bonded iodine, the distances are 2.994(4)~{\AA} for Nb$_4$Se$_4$I$_4$,\cite{Nb4Se4I4} and a range from 2.7459(10)~{\AA} to 2.9428(9)~{\AA} for Nb$_2$Se$_2$I$_6$.\cite{Nb2Se2I6} In these two compounds, all iodine ions are directly bonded to niobium.
On the contrary, the closest Nb--I distance is 4.839(2)~{\AA} for (NbSe$_4$)$_3$I,\cite{NS3I_first_structure} and 4.8273(5)~{\AA} for (NbSe$_4$)$_{3.33}$I,\cite{NS333I_structure} where all the iodine ions are interchain.
The Nb--I distances for all these compounds are plotted in Figure \ref{fig: distance}, to emphasize that half of the iodine ions in {\nsii} follow the expected trends for direct intrachain bonding, while the other half are weakly associated interchain ions.

In each formula unit, fourteen selenium ions are dimerized, having a 1- valence per Se, while the singular selenium ion has a 2- valence.
There are half of niobium ions bonded to iodine, and the other half unbonded, but the valence of all the niobium ions is 5+, which suggests {\nsii} may be an insulator, because there are no free $d$ electrons, unlike (TaSe$_4$)$_2$I.
{\nsii} has a space group of $P2_1/c$ (No. 14), with a much lower symmetry than (TaSe$_4$)$_2$I (space group $I422$, No. 97), due to the divalent selenium and two types of iodine ions.
However, one commonality between {\nsii} and (TaSe$_4$)$_2$I is that they both have chiral chains.
In \nsii, each layer of chains stacked along the [010] direction shares the same chirality, with the screw direction opposed on each successive (010) layer, leading to a material that has no net chirality, confirmed by the $c$-glide in space group $P2_1/c$.

\begin{figure}
    \centering
    \includegraphics[width=\columnwidth]{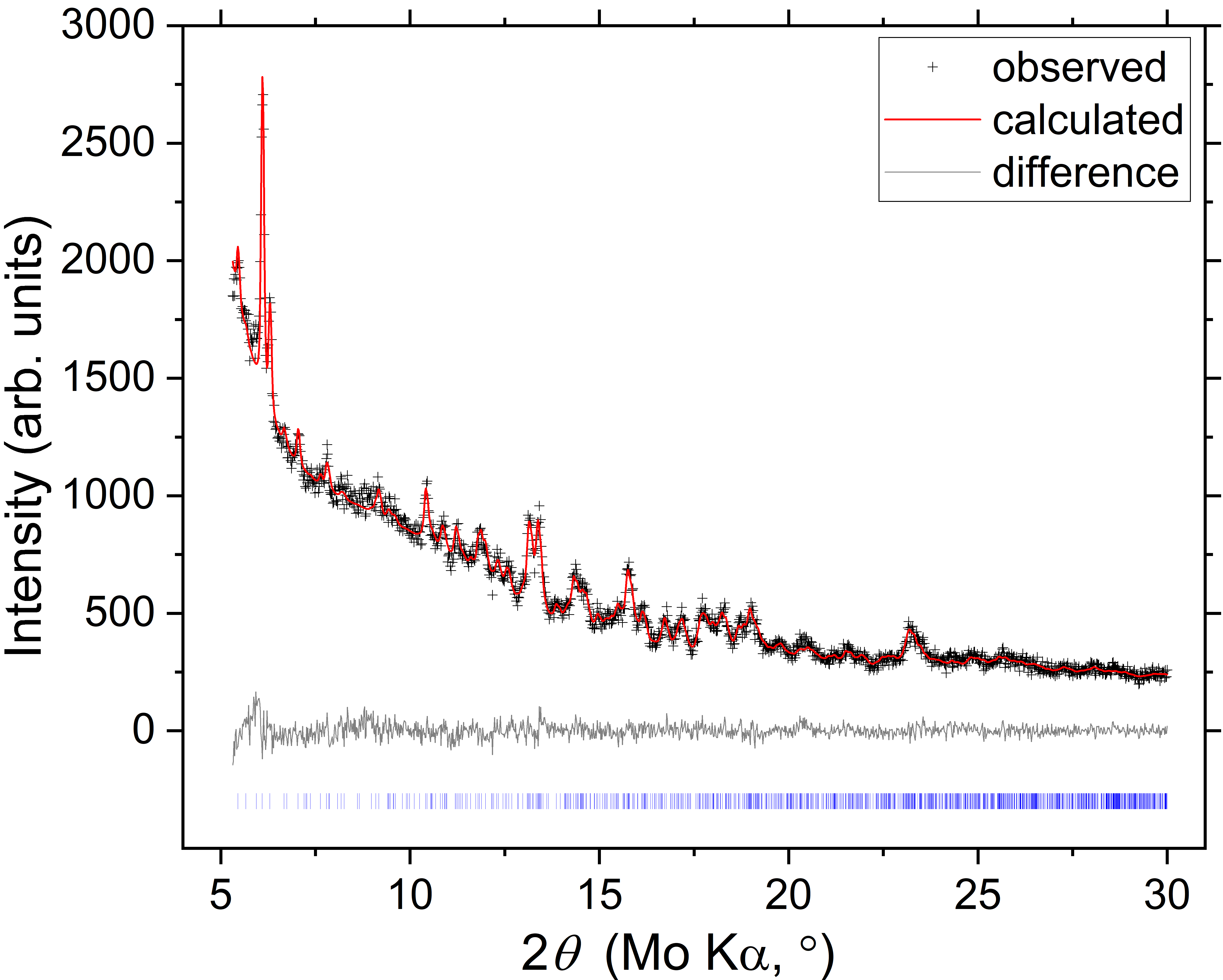}
    \caption{Rietveld refinement to the powder XRD pattern of {\nsii}.}
    \label{fig: XRD}
\end{figure}

Powder XRD was carried out to verify the refined crystal structure and the phase purity of the crystals. The Rietveld refined result is shown in Figure \ref{fig: XRD}. Most peaks are weak due to the difficulty of grinding the crystals (see Methods section) and the existence of heavy elements and thus high X-ray absorption. All the peaks matched the calculated positions and there is no amorphous background, thus {\nsii} crystal structure is confirmed and the crystals can be considered pure.

\subsection{Band structure calculations}

\begin{figure}
    \centering
    \includegraphics[width=\columnwidth]{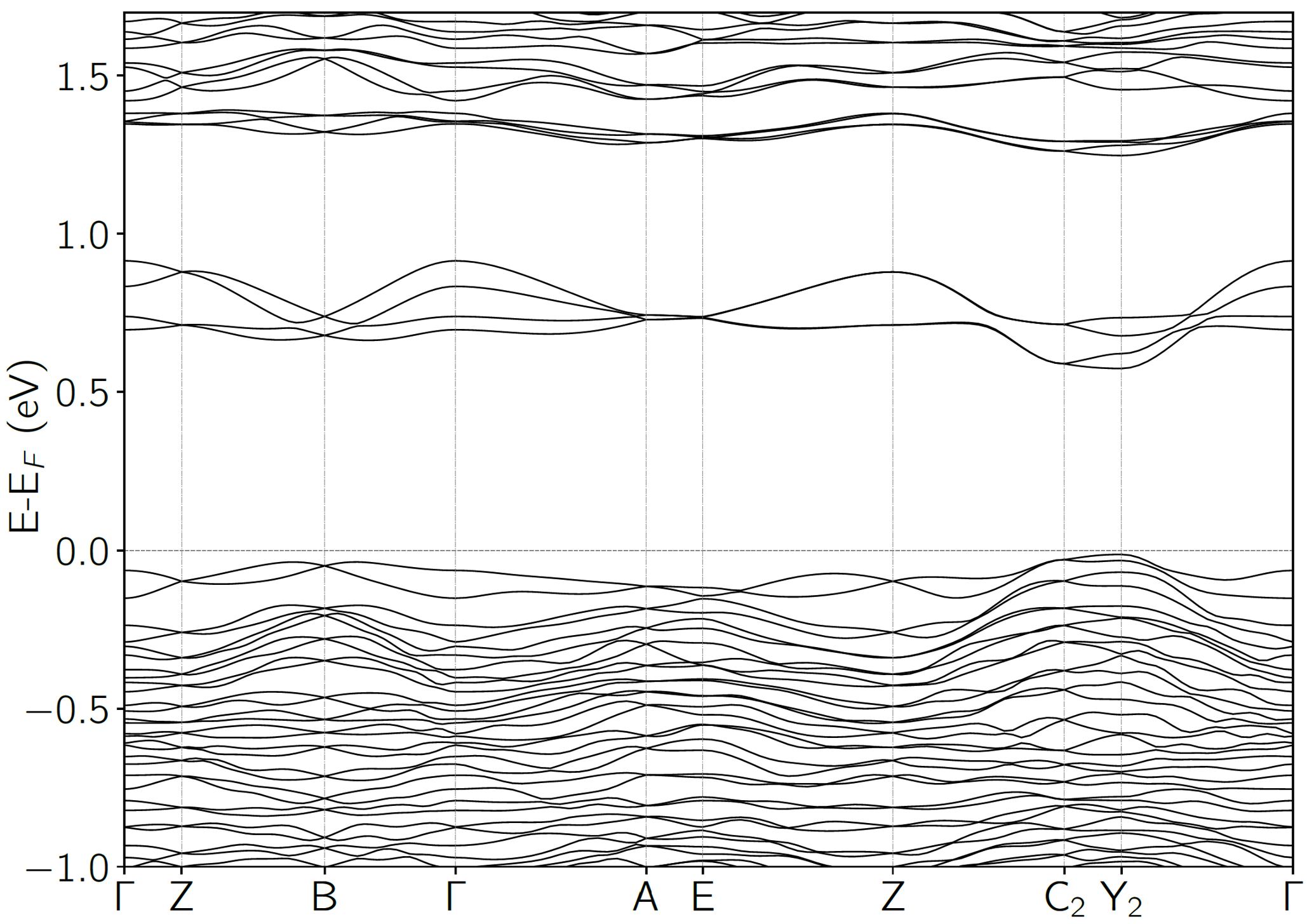}
    \caption{DFT-PBE calculated electronic band structure of {\nsii}, showing band insulator behavior. The direct band gap at Y$_2$ is 0.59~eV.}
    \label{fig: bands}
\end{figure}

The band structure of {\nsii} calculated by DFT-PBE is shown in Figure \ref{fig: bands}, while the DFT-mBJ result is included in Figure S2 as a comparison with PBE.\cite{supplement} DFT calculations with both approximations predict {\nsii} to be an semiconductor with a direct band gap. For the PBE approximation, the band gap is direct at Y$_2$ and $E_\mathrm{g} = 0.59$~eV while the direct band gap for mBJ method is also at Y$_2$ and $E_\mathrm{g}=0.68$~eV. Both approximations give very similar results.

\begin{figure*}
    \centering
    \includegraphics[width=\textwidth]{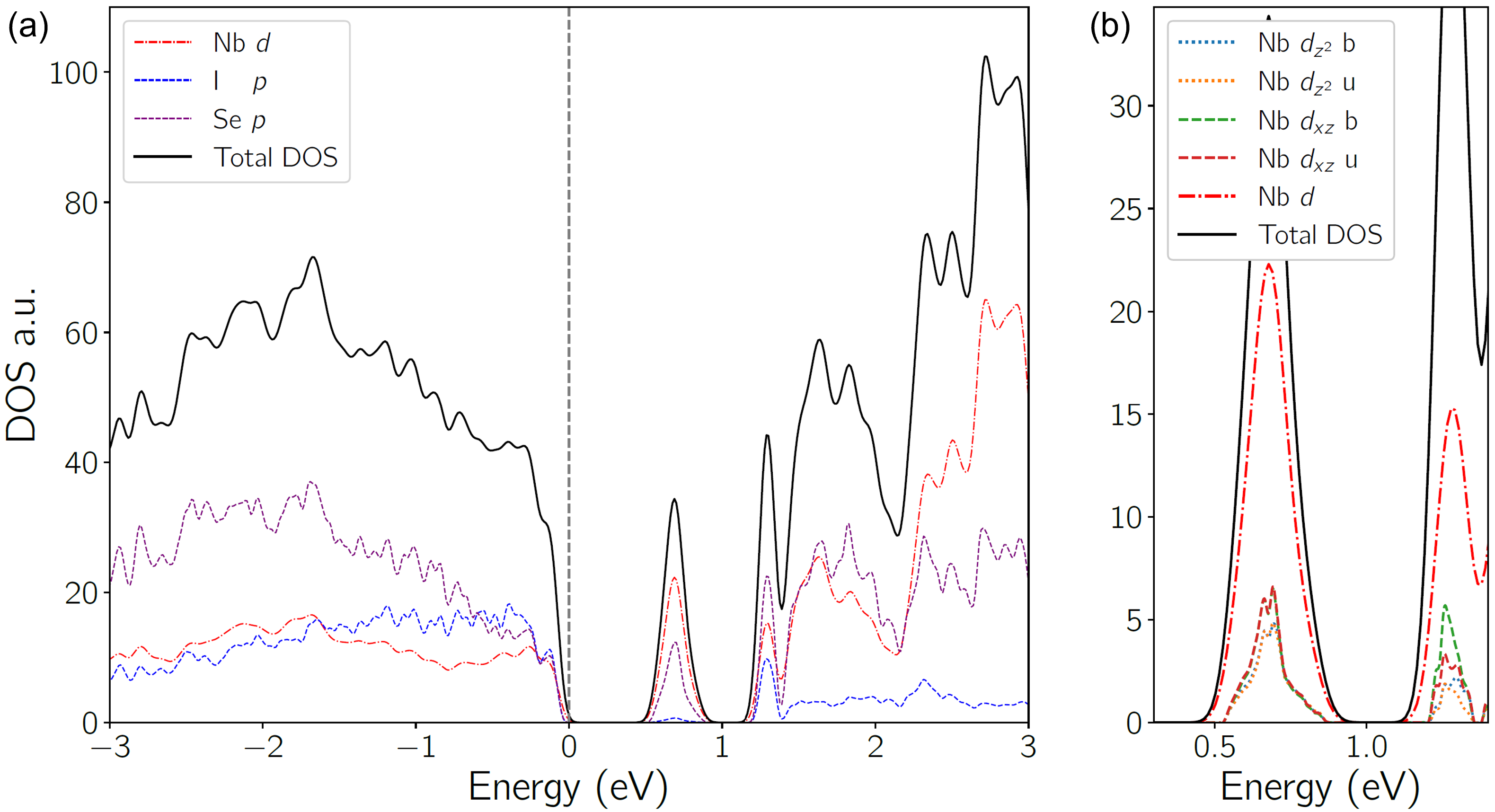}
    \caption{(a) Density of states of {\nsii}, decomposed on different types of atoms. (b) The DOS is further decomposed on Nb $d$ orbitals, separating I-bonded (Nb b) and I-unbonded (Nb u) niobium ions.}
    \label{fig: DOS}
\end{figure*}

The most notable feature of this band structure is an isolated group of 8 bands that comprise the conduction band minimum, which combine to form an isolated peak in the electronic density of states (DOS), shown in Figure \ref{fig: DOS}. The character of these bands are mainly Nb $d$ orbitals and Se $p$ orbitals. To differentiate \nsii\ from (TaSe$_4$)$_2$I and other compounds with mixed valence, we can separate Nb ions into those bonded to I$^-$ (Nb b) and those unbonded to I$^-$ (Nb u), and compare their contributions to the DOS in Figure \ref{fig: DOS}(b), with a focus just above $E_{\mathrm{F}}$. Figure S3 in Supporting Information shows the DOS contributions from bonded and unbonded I$^-$ in the whole range, and they are essentially identical.\cite{supplement}

The main Nb contribution to the set of 8 conduction bands comes from $d_{xz}$ and $d_{z^2}$ orbitals. 
While strictly linear chains, such as those in (TaSe$_4$)$_2$I, have orbital occupation near $E_{\mathrm{F}}$ of $dz^2$ only,\cite{gressier1984electronic} \nsii\ chains have a sideways step every fourth Nb, which necessitates occupation of another orbital.
Both sets of niobium (Nb-b bonded to I$^-$, and Nb-u bonded to (Se$_2$)$^2-$ only) are electronically equivalent around the valence band maximum and the 8 low-energy conduction bands, with some minor differences around 1.2~eV above $E_{\mathrm{F}}$.
This similarity confirms that charge is not highly modulated along the chain, and the assignment of Nb$^{5+}$ is consistent throughout the compound. Without mixed valence on Nb, the Nb $4d$ orbitals are unoccupied, far from the Fermi energy, and there is no conducting state that would be expected to lead to a CDW.

\subsection{Phase transition determination}

\begin{figure}
    \centering
    \includegraphics[width=\columnwidth]{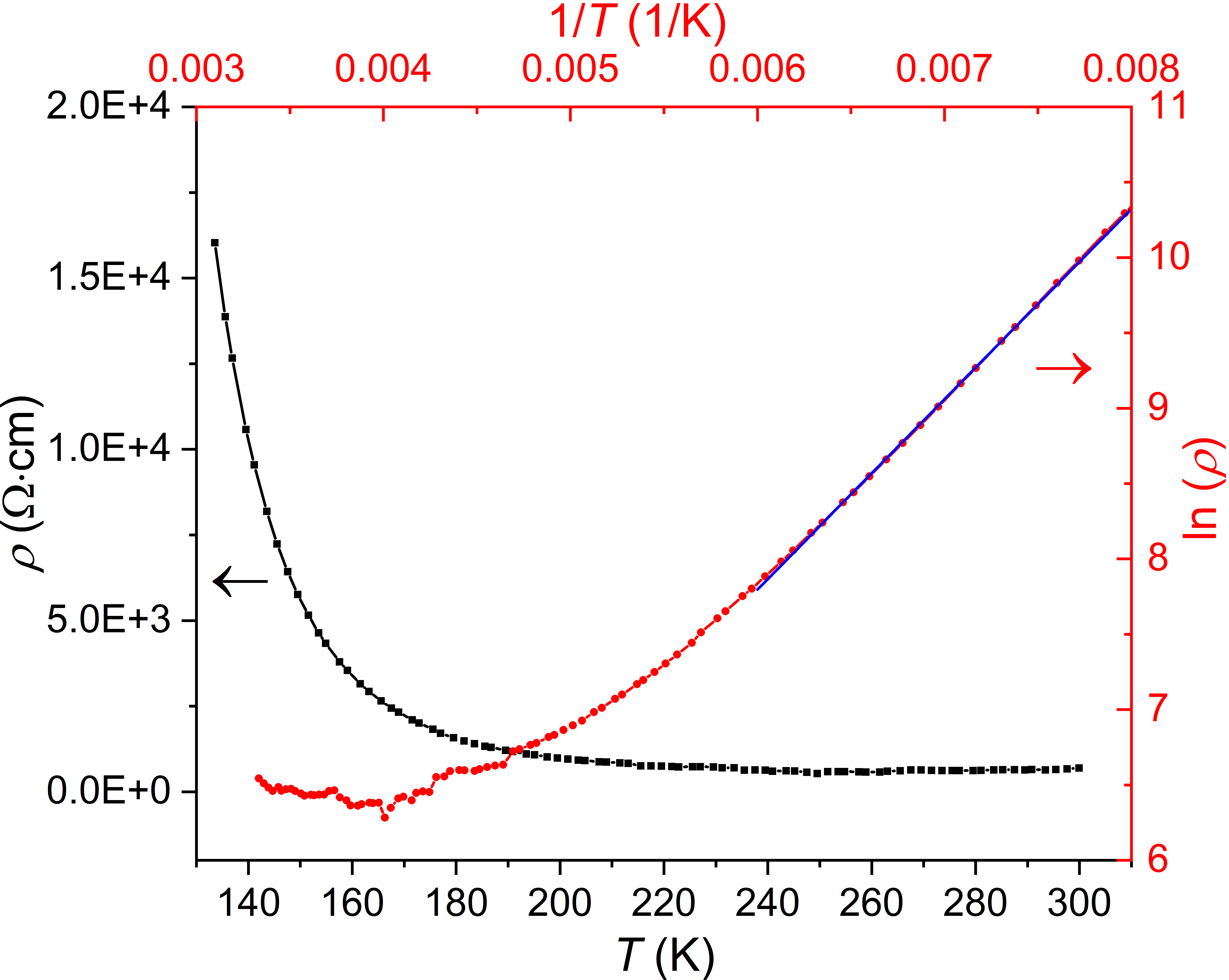}
    \caption{Four-point resistivity of a bulk {\nsii} crystal. The black and red dots and lines show $\rho$ vs. $T$ and ln ($\rho$) vs. $1/T$, respectively. The blue straight line fits part of the ln ($\rho$) vs. $1/T$ curve, and Arrhenius equation was fitted to determine the activation energy $E_\mathrm{a}$.}
    \label{fig: resistivity}
\end{figure}

In order to see if {\nsii} undergoes a CDW phase transition, like (TaSe$_4$)$_2$I does, an in-line four-point resistance measurement was done on a large bulk needle-shaped {\nsii} crystal, and the result is shown in Figure \ref{fig: resistivity}.
The resistivity increases with decreasing temperature, until the resistance exceeds the range of the measurement.
The ln ($\rho$) versus $1/T$ curve was also plotted in red, and the empirical Arrhenius fitting $\rho={\rho}_0\mathrm{exp}(E_\mathrm{a}/{k_{\mathrm{B}}T})$ was done for the linear part of the curve. Here $E_\mathrm{a}$ is the activation energy, which can be smaller than the band gap $E_\mathrm{g}$, because the material may contain impurities which have electron energy levels in the band gap.
The calculated activation energy is $E_\mathrm{a}=0.1$~eV, much smaller than the DFT calculated band gap $E_\mathrm{g}=0.6$~eV.
From resistivity measurement, {\nsii} behaves like a normal semiconductor upon cooling from 300~K to 140~K. No sign of an anomaly has been observed.

As a comparison, a nano-device was made on an exfoliated {\nsii} crystal (see Methods section) and the result is shown in Figure S4 in Supporting Information.\cite{supplement} Only a small portion of the data is shown because the nano-device became highly resistive at low temperature and the silicon substrate contributed to the resistance measured. The activation energy determined is 0.06~eV, close to that of the bulk crystal.

Room temperature and cryogenic temperature XRD was performed on a single crystal {\nsii}, and the result is shown in Figure S5 (290~K) and S6 (8.2~K) in Supporting Information.\cite{supplement} The diffraction pattern did not change as the sample was warmed up from 8.2~K to room temperature, nor did satellite peaks form. This demonstrated there is no structural transition between room temperature and 8.2~K.

Finally, differential scanning calorimetry (DSC) was performed on {\nsii}, with the results shown in Figure S7 in Supporting Information.\cite{supplement}
No transition peaks can be seen in the range of -125$^\circ$C to 25$^\circ$C.

The resistivity measurements on bulk single crystal and nano-device, the single crystal XRD at cryogenic temperature, and the DSC measurements did not show a phase transition such as the CDW. Taking the band structure calculated by DFT into consideration, as well as the fact that all Nb ions are charge-balanced, it can be concluded that unlike (TaSe$_4$)$_2$I, the new material {\nsii} is a semiconductor with a medium band gap, and it stays as a normal semiconductor.

\section{Conclusions}

Single crystals of a new transition metal chalcogenide {\nsii} were grown by CVT and the crystal structure was solved. DFT predicts {\nsii} to be a semiconductor with a band gap of around 0.6~eV. Resistivity measurements showed the Arrhenius activation energy to be 0.1~eV, likely signifying shallow impurity energy levels. Low-temperature XRD and DSC did not show evidence of any phase transformation, and {\nsii} demonstrated itself to be a band insulator with a moderate band gap. It remains to be seen if {\nsii} can have its band structure engineered by chemical doping, strong field or external strain.
Due to the complexities of transport measurements carried out on related materials such as (TaSe$_4$)$_2$I, it is important to verify how differences in electronic structure or symmetry might allow or preclude the observed magnetoresistance data that is thought to arise from Weyl or axion behavior. Furthermore, the structure types and motifs in this class of materials are still being discovered, which may yet result in more exotic transport phenomena than have yet been proposed. 

\section{Supporting Information}

The Supporting Information is available free of charge, containing: an image of reaction tubes after CVT, a table with atomic positions of \nsii\, DFT-PBE and DFT-mBJ band structure calculations, DOS of I-bonded and I-unbonded Nb in the whole DFT range, resistivity measurement of a nano-device, single crystal XRD at 290~K and 8.2~K with indexing, and differential scanning calorimetry.

\section{Acknowledgments}

Crystal growth, transport, and microstructure characterization were supported by the Center for Quantum Sensing and Quantum Materials, an Energy Frontier Research Center funded by the U. S. Department of Energy, Office of Science, Basic Energy Sciences under Award DE-SC0021238. The authors acknowledge the use of microscopy facilities at the Materials Research Laboratory Central Research Facilities, University of Illinois, partially supported by NSF through the University of Illinois Materials Research Science and Engineering Center DMR-1720633. S.B. acknowledges support through the Early Postdoc Mobility Fellowship from the Swiss National Science Foundation (Grant number P2EZP2 191885). F. J. acknowledges funding from the Spanish MCI/AEI/FEDER (grant PID2021-128760NB-I00). M. G. V. acknowledges the Spanish Ministerio de Ciencia e Innovaci\'{o}n (grant PID2019- 109905GB-C21) and the Deutsche Forschungsgemeinschaft (DFG, German Research Foundation) GA 3314/1-1 – FOR 5249 (QUAST). P.A. acknowledges support from the EPiQS program of the Gordon and Betty Moore Foundation, grant no. GBMF9452.

%



\clearpage

\bibliography{nsii}

\clearpage 



\begin{tocentry}
\centering
\includegraphics[height=4.45cm]{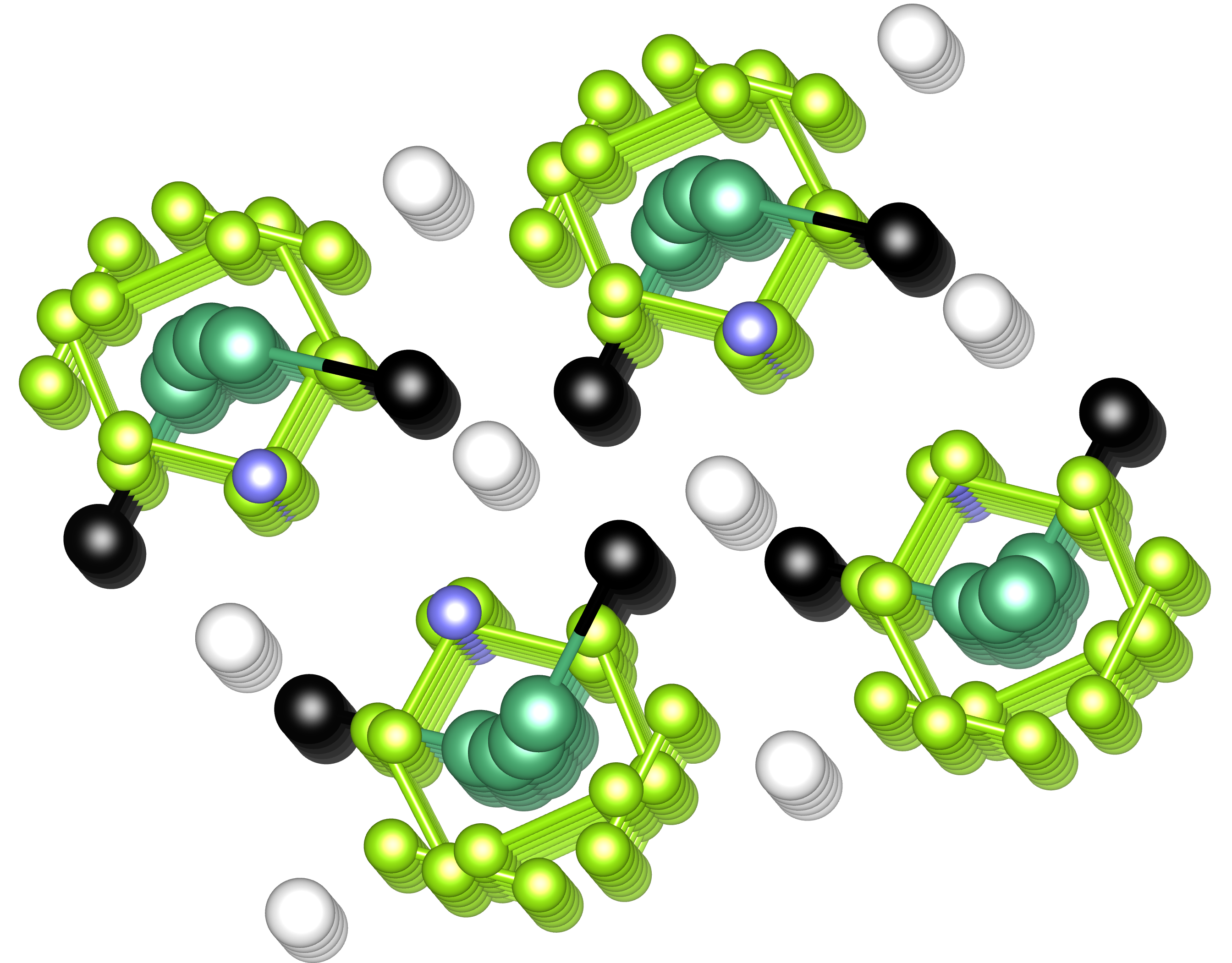}\\

\end{tocentry}

\end{document}


\renewcommand{\thetable}{S\arabic{table}}%
\renewcommand{\thefigure}{S\arabic{figure}}%

\newcommand{\dps}[1]{\textbf{\textcolor{blue}{DPS: #1}}}
\newcommand{\oq}[1]{\textbf{\textcolor{red}{OQ: #1}}}


\begin{center}
\Large 
\textbf{A new quasi-one-dimensional transition metal chalcogenide semiconductor (Nb$_4$Se$_{15}$I$_2$)I$_2$}\\
\vspace{1em}
Supporting Information\\
\vspace{1em}
\normalsize
Kejian Qu, Zachary W. Riedel, Iri\'{a}n S\'{a}nchez-Ram\'{i}rez, Simon Bettler, Junseok Oh, Emily N. Waite, Nadya Mason, Peter Abbamonte, Fernando de Juan Sanz, Maia G. Vergniory, Daniel P. Shoemaker

\vspace{3em}

\Large 
\textbf{Contents}

\end{center}

\begin{enumerate}
  \item Figure S1: picture of tubes after reaction.
  \item Table S1: atomic positions of (Nb$_4$Se$_{15}$I$_2$)I$_2$.
  \item Figure S2: DFT-PBE and DFT-MBJ calculations of (Nb$_4$Se$_{15}$I$_2$)I$_2$.
  \item Figure S3: DOS of I-bonded and I-unbonded Nb in the whole DFT range.
  \item Figure S4: resistance vs. temperature for (Nb$_4$Se$_{15}$I$_2$)I$_2$ nano-device.
  \item Figure S5: single crystal XRD at 290~K.
  \item Figure S6: single crystal XRD at 8.2~K with indexing.
  \item Figure S7: DSC result from -125$^\circ$C to 25$^\circ$C.
\end{enumerate}

\pagebreak

\begin{figure}[h]
    \vspace{10em}
    \centering
    \includegraphics[width=\columnwidth]{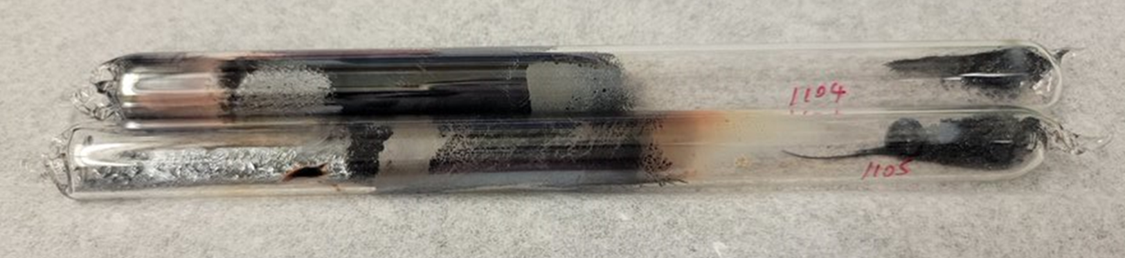}
    \caption{Picture of the sealed tubes after CVT reaction. For the top tube, the starting ingredients were loaded to the right side, which was kept 90 hours at 420$^\circ$C and the low temperature side was at 280$^\circ$C. After reaction, the lower temperature half of the tube was coated by iodine or iodine containing compound, where (Nb$_4$Se$_{15}$I$_2$)I$_2$ crystals grew. The lower tube had a high/low temperature set-points of 380$^\circ$C/260$^\circ$C, with the same heating and holding time. The crystals were smaller and the yield decreased as well, and the low-temperature end of the tube did not have crystals. This demonstrates the sensitivity of (Nb$_4$Se$_{15}$I$_2$)I$_2$ crystal growth to the reacting temperature.}
    \label{fig: tube}
\end{figure}

\begin{table}[h]
    \centering
    \caption{Atomic positions of (Nb$_4$Se$_{15}$I$_2$)I$_2$.}
    \begin{tabular}{c c c c c}
    \hline
Atomic site label & x-fraction & y-fraction & z-fraction & $U_{\mathrm{iso}}$ \\
    \hline
I-001 & 0.00110(4) & 0.41817(2) & 0.18602(3) & 0.022934 \\
I-002 & 0.22240(4) & 0.02954(2) & 0.01402(3) & 0.029585 \\
I-003 & 0.34201(5) & 0.04986(2) & 0.35662(3) & 0.034316 \\
I-004 & 0.71913(4) & 0.30544(2) & 0.03499(4) & 0.036386 \\
Nb-05 & 0.17103(5) & 0.63827(2) & 0.37801(4) & 0.015416 \\
Nb-06 & 0.09724(5) & 0.12563(2) & 0.35140(4) & 0.016340 \\
Nb-07 & 0.28573(5) & 0.36702(2) & 0.17245(4) & 0.014715 \\
Nb-08 & 0.56191(5) & 0.35955(2) & 0.40002(4) & 0.015090 \\
Se-09 & 0.16947(6) & 0.44501(2) & 0.02495(4) & 0.019974 \\
Se-0A & 0.08474(5) & 0.30633(2) & 0.02190(4) & 0.017271 \\
Se-0B & 0.33055(5) & 0.18206(2) & 0.48127(4) & 0.019589 \\
Se-0C & 0.17427(5) & 0.58645(2) & 0.56707(4) & 0.019298 \\
Se-0D & 0.16562(6) & 0.82051(2) & 0.08956(4) & 0.021012 \\
Se-0E & 0.42460(6) & 0.44538(2) & 0.30836(4) & 0.020022 \\
Se-0F & 0.53657(6) & 0.10665(2) & 0.10266(4) & 0.021851 \\
Se-0G & 0.26262(6) & 0.32654(2) & 0.36910(4) & 0.021728 \\
Se-0H & 0.42410(6) & 0.27981(2) & 0.27562(4) & 0.021657 \\
Se-0I & 0.58813(6) & 0.19755(2) & 0.09824(4) & 0.021008 \\
Se-0J & 0.18781(6) & 0.61401(2) & 0.15092(4) & 0.021096 \\
Se-0K & 0.55874(5) & 0.40015(2) & 0.18379(4) & 0.020027 \\
Se-0L & 0.01473(6) & 0.21251(2) & 0.22439(4) & 0.022028 \\
Se-0M & 0.14043(6) & 0.14598(2) & 0.14515(5) & 0.024782 \\
Se-0N & 0.09210(7) & 0.55154(2) & 0.26367(5) & 0.025453 \\
    \hline
    \end{tabular}
    \label{tab: atomic positions}
\end{table}

\begin{figure}[h]
    \centering
    \includegraphics[width=0.95\columnwidth]{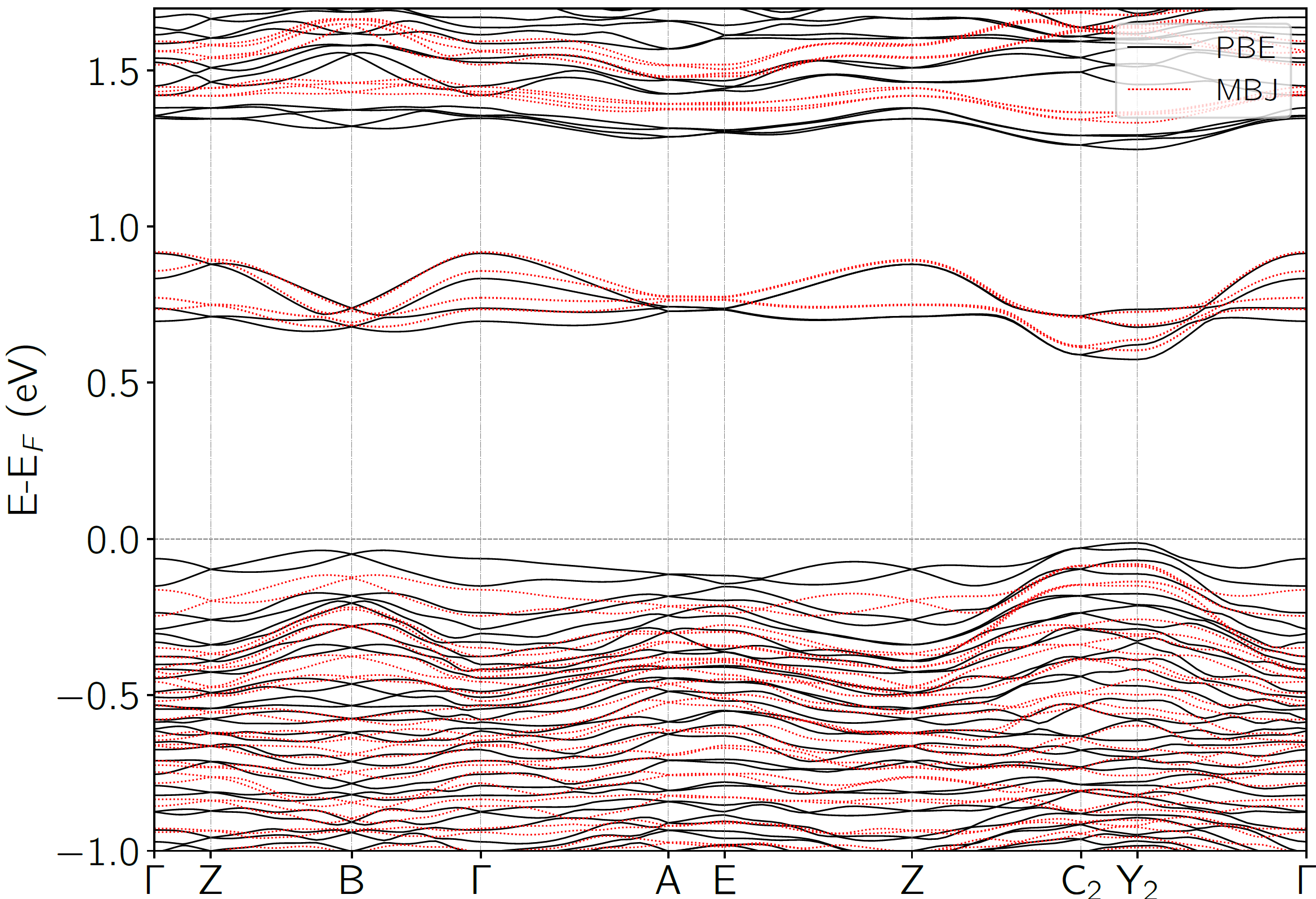}
    \caption{DFT-PBE and DFT-mBJ calculated electronic band structure of (Nb$_4$Se$_{15}$I$_2$)I$_2$, showing band insulator behavior. The direct band gap at Y$_2$ is 0.59~eV and 0.68~eV for PBE and mBJ approximations, respectively.}
    \label{fig: PBEMBJ}
\end{figure}

\begin{figure}[h]
    \centering
    \includegraphics[width=0.95\columnwidth]{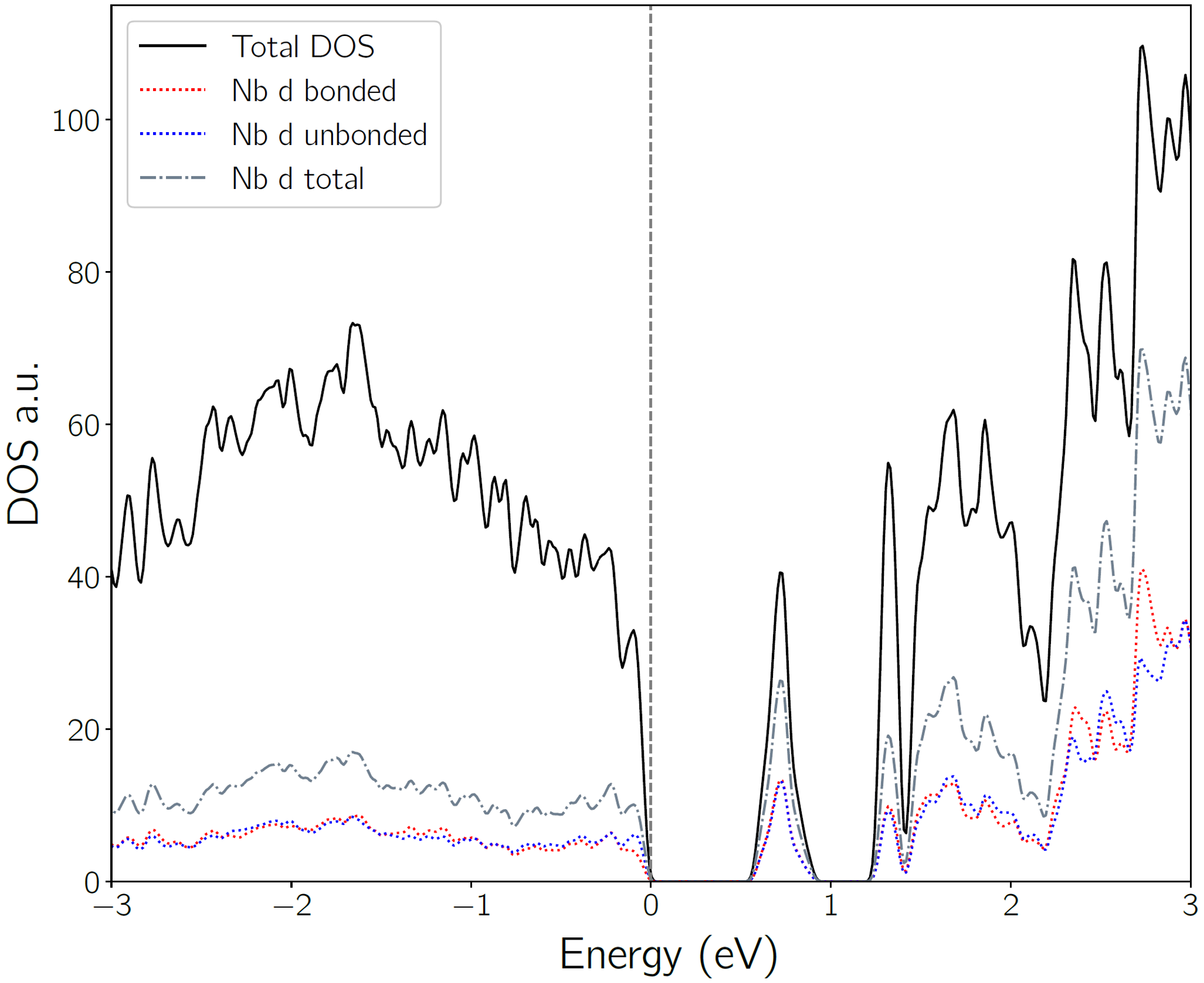}
    \caption{Density of states of I-bonded (Nb b) and I-unbonded (Nb u) niobium ions in the whole range.}
    \label{fig: DOS whole range}
\end{figure}

\begin{figure}[h]
    \centering
    \includegraphics[width=0.75\columnwidth]{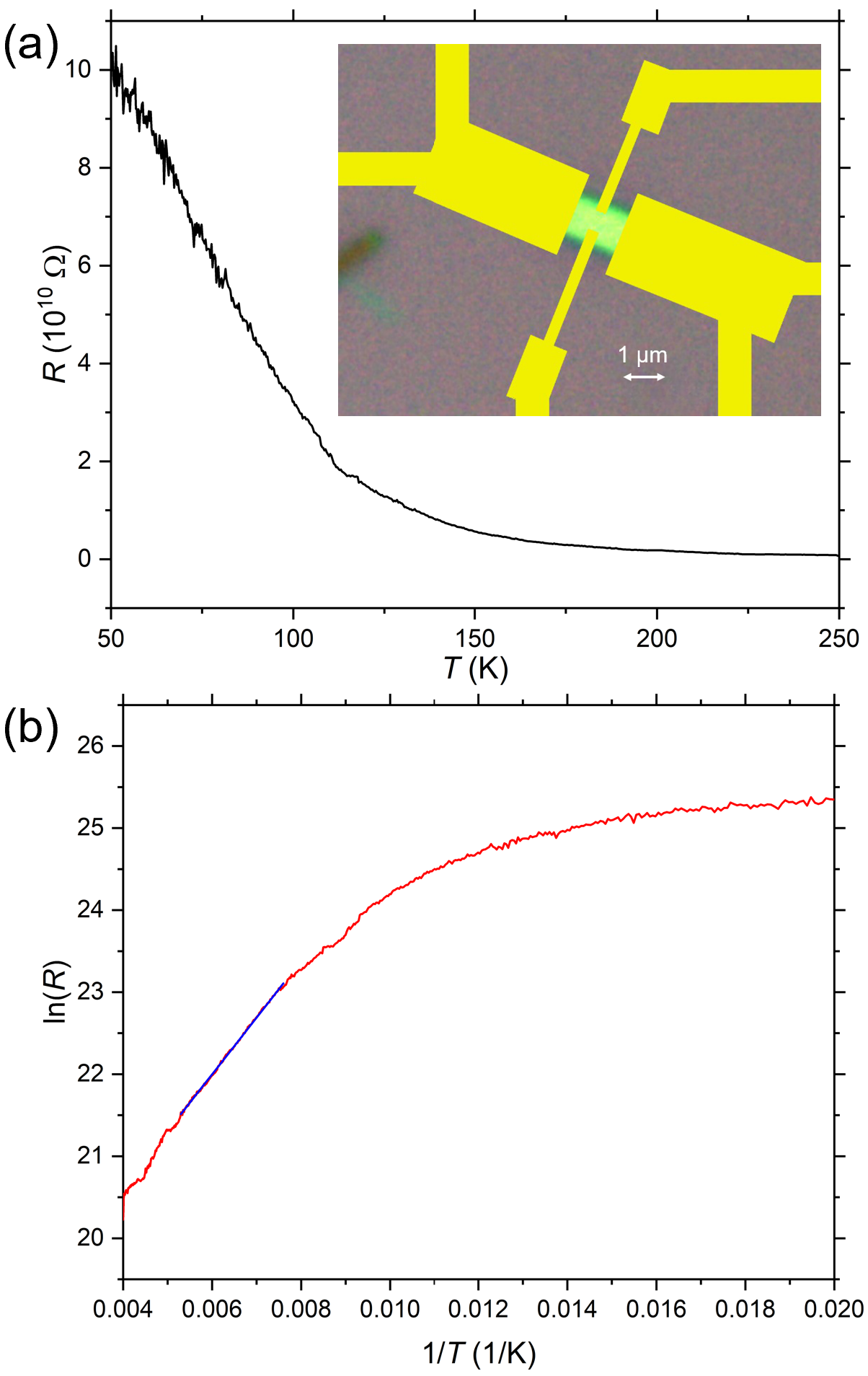}
    \caption{(a) $R$ vs. $T$ for an exfoliated nano-device, where the inset shows the microscope picture of the device in green, with yellow leads for measurement. The sample became highly resistive at low temperatures, and the data became very noisy. (b) ln ($R$) vs. $1/T$ curve, where the blue line is the straight-line part used for Arrhenius equation fitting. The activation energy fitted is 0.06~eV, close to the result from bulk crystal.}
    \label{fig: Junseok}
\end{figure}

\begin{figure}[h]
    \centering
    \includegraphics[width=0.9\columnwidth]{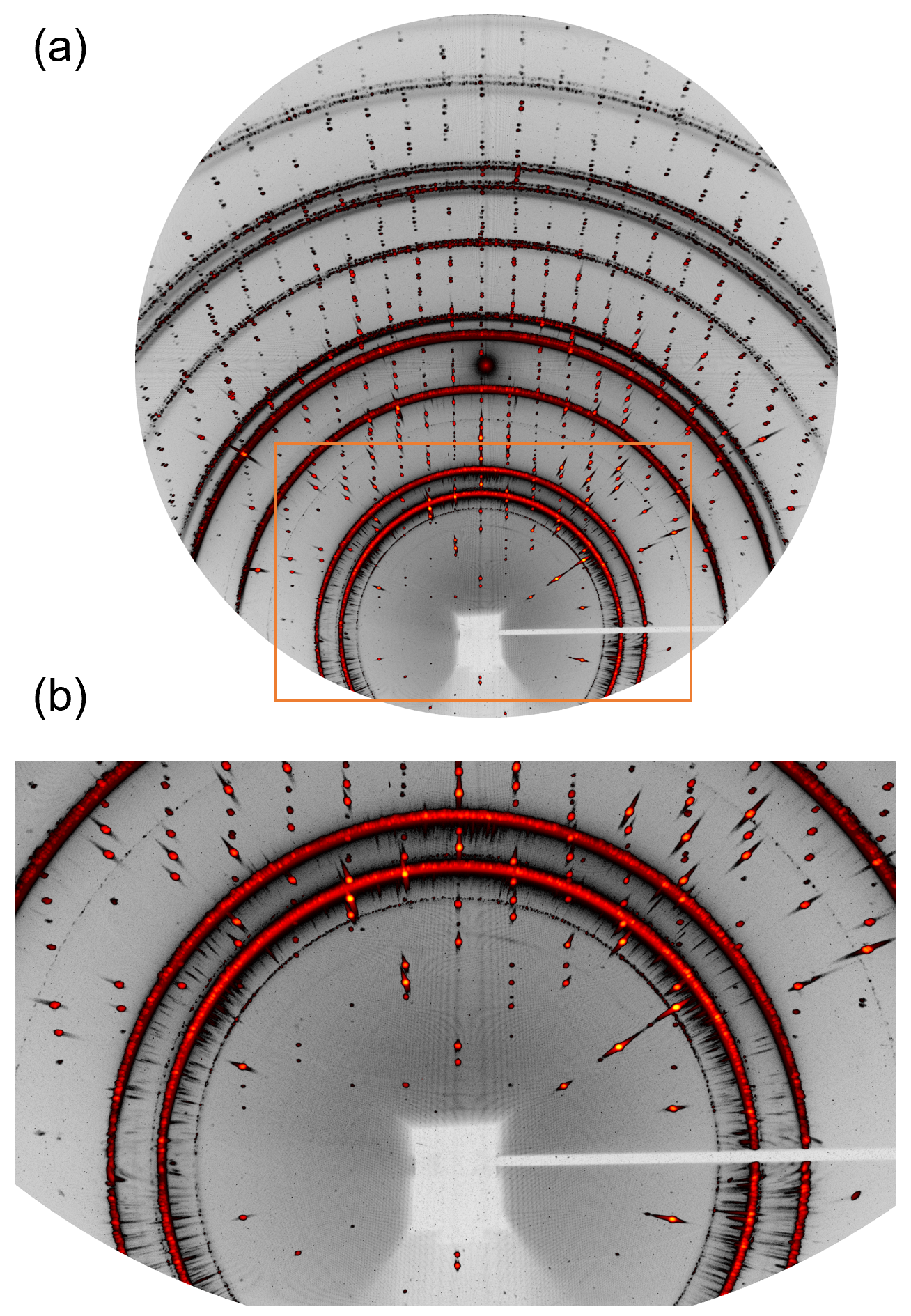}
    \caption{(a) Single crystal XRD for (Nb$_4$Se$_{15}$I$_2$)I$_2$ at 290~K. (b) Enlarged view of the region in the rectangular box in (a).}
    \label{fig: lowTXRD}
\end{figure}

\begin{figure}[h]
    \centering
    \includegraphics[width=0.9\columnwidth]{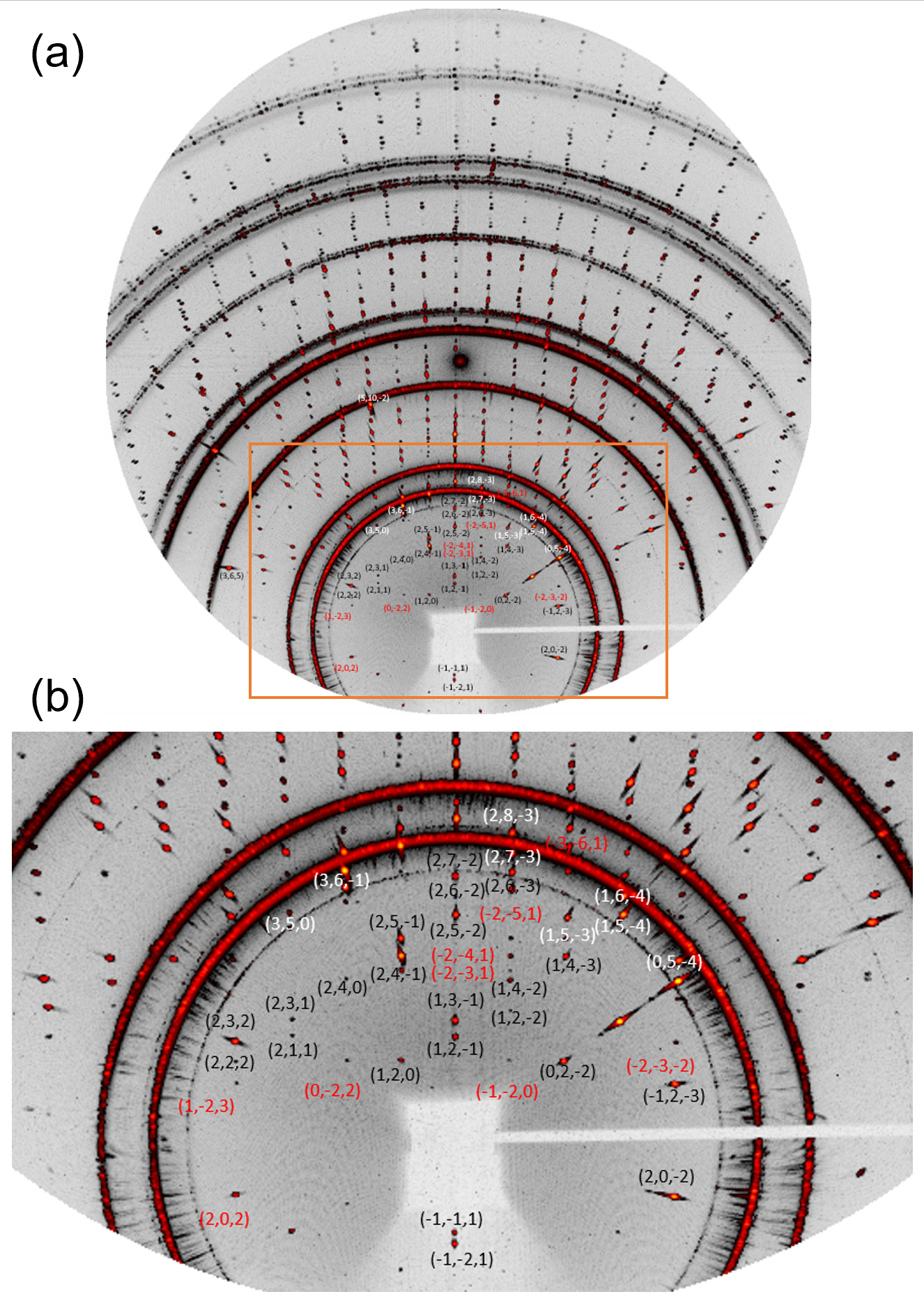}
    \caption{(a) Single crystal XRD for (Nb$_4$Se$_{15}$I$_2$)I$_2$ at 8.2~K. (b) Enlarged view of the region in the rectangular box in (a), with peaks indexed. No extra reflections were seen to arise from any symmetry-lowering structural transition compared to Figure S5.}
    \label{fig: lowTXRD}
\end{figure}

\begin{figure}[h]
    \centering
    \includegraphics[width=0.95\columnwidth]{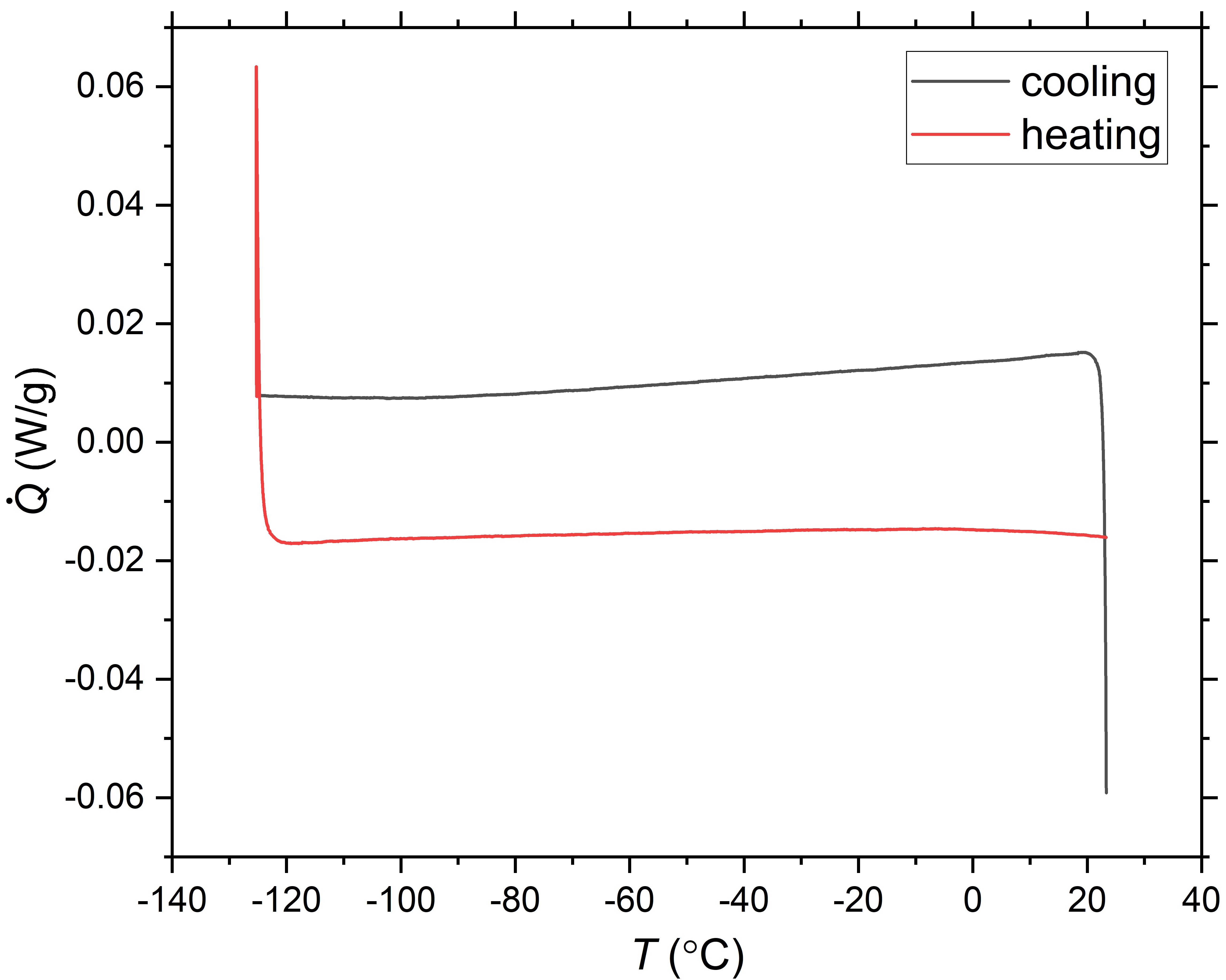}
    \caption{Differential scanning calorimetry measurement for (Nb$_4$Se$_{15}$I$_2$)I$_2$ from -125$^\circ$C to 25$^\circ$C. The sample was measured in two cooling and heating cycles, and only the second cycle is shown in the figure. No transition peak can be seen.}
    \label{fig: DSC}
\end{figure}
